\documentclass[12pt]{article}

\newlength{\abstractwidth}
\renewcommand{\thanks}[1]{\footnote{#1}} % Use this for footnotes

\makeatletter
\newcounter{fig}
\renewcommand\thefig{\arabic{fig}}

\def\fps@fig{tbp}
\def\ftype@fig{1}
\def\ext@fig{lof}
\def\fnum@fig{\figurename~\thefig}

\newenvironment{fig*}
               {\@dblfloat{fig}}
               {\end@dblfloat}

\makeatother
\usepackage{epsfig}

\newcommand{\be}{\begin{equation}}
\newcommand{\bea}{\begin{eqnarray}}
\newcommand{\eea}{\end{eqnarray}}
\newcommand{\ee}{\end{equation}}

\def\ba{\begin{eqnarray}}
\def\ea{\end{eqnarray}}

\def\eg{{\it e.g.~}}
\def\ie{{\it i.e.~}}
\def\nn{\nonumber}

\def\tr{{\rm tr}}

\def\half{ {1\over 2}}

\def\unit{1 \hskip-.3em \raise2pt\hbox{$ \scriptstyle |$ } }

\def\lab{\label}

\def\oj{\overline{\j}}
\def\oq{\overline{\q}}
\def\of{\overline{\f}}

\def\mb{\overline{m}_0}
\def\sH{H\!\!\!\!/}
\def\sF{F\!\!\!\!/}

\newcommand{\6}{\partial}

\newcommand{\eps}{\epsilon}

\newcommand{\rc}{\nonumber\\}
\def\sym{${\cal N}=1$ SYM }

\def\half{\frac12}

\def\le{\left}
\def\ri{\right}

\def\mng{${\rm MN}_{\g}$ }
\def\a{\alpha}
\def\b{\beta}

\def\d{\delta}
\def\e{\epsilon}           % Also, \varepsilon
\def\f{\varphi}               %      \varphi
  
\def\g{\gamma}

\def\j{\psi}
\def\k{\kappa}                    % Also, \varkappa (see below)
\def\l{\lambda}
\def\m{\mu}
\def\n{\nu}
  
\def\q{\theta}                    %     \vartheta
\def\qt{\tilde{\theta}}
\def\ft{\tilde{\varphi}}
\def\gt{\tilde{\gamma}}
\def\tht{\tilde{\theta}}
\def\gb{\bar{\gamma}}
\def\r{\rho}                                     %     \varrho
\def\s{\sigma}                                   %     \varsigma
\def\t{\tau}

\def\W{\Omega}

\def\ca{{\cal A}}

\def\cz{{\cal Z}}

\def\bop#1{\setbox0=\hbox{$#1M$}\mkern1.5mu
        \vbox{\hrule height0pt depth.04\ht0
        \hbox{\vrule width.04\ht0 height.9\ht0 \kern.9\ht0
        \vrule width.04\ht0}\hrule height.04\ht0}\mkern1.5mu}
\def\leftrighthookfill#1{$\mathsurround=0pt \mathord\hook#1
       \hrulefill\mathord\hook#1$}
\def\underhook#1{\vtop{\ialign{##\crcr                 % |_| under
       $\hfil\displaystyle{#1}\hfil$\crcr
       \noalign{\kern-1pt\nointerlineskip\vskip2pt}
       \leftrighthookfill5\crcr}}}

\def\to{\rightarrow}

\makeatletter

\begin{document}

\begin{titlepage}

\vskip -.8cm

\rightline{\small{\tt CPHT RR-013.0206}}

\begin{center}

\vskip 2.5 cm

{\LARGE\bf Probing Universality in the Gravity Duals of\\[5pt] 
\sym by $\gamma$ Deformations}

%\vskip 1.5cm
%{\large }
%\vskip 1.2cm

%\vskip 0.5cm

\vskip 2.cm

{\large Umut G{\"u}rsoy}

\vskip 0.6cm

{\it {Centre de Physique Th\'eorique\\ \'Ecole Polytechnique\\
91128 Palaiseau, France} \\
\vskip 0.4 cm

E-mail: {\tt gursoy@cpht.polytechnique.fr}} \\

\end{center}

\vspace{1.9cm}

\begin{center}
{\bf Abstract}
\end{center}

Recently, a one-parameter deformation of the Maldacena-N\'u\~nez 
dual of ${\cal N}=1$ SYM theory was constructed in hep-th/0505100. 
According to the Lunin-Maldacena conjecture, 
the background is dual to pure \sym in the IR coupled to a 
KK sector whose dynamics is altered by a dipole deformation 
that is proportional to the deformation parameter 
$\gamma$. Thus, the deformation serves to 
identify the aspects of the gravity backgrounds that bear 
the effects of the KK sector, hence non-universal in the dual gauge theory. 
We make this idea concrete by studying a Penrose limit 
of the deformed MN theory. We obtain an exactly solvable 
pp-wave that is conjectured to describe the IR dynamics of KK-hadrons in the
field theory. The spectrum, the thermal partition function 
and the Hagedorn temperature are calculated. 
Interestingly, the Hagedorn temperature turns out to be independent 
of the deformation parameter. 

\noindent

\end{titlepage}

\newpage

\section{Introduction}
\setcounter{equation}{0}

One fundamental obstacle in the holographic approach\cite{adscft} 
to Yang-Mills theory is the entanglement of pure YM dynamics 
with the dynamics of the Kaluza-Klein sectors that arise 
from compactificaton of the dual 10D geometry to 4 dimensions\cite{Witten:1998qj}.   
Generically, the supergravity approximation becomes invalid 
in the limit where the KK sector becomes very massive 
hence decouples from the pure YM dynamics. 
Thus, one should go beyond the supergravity approximation 
and study the full string theory on the dual 10D backgrounds. 
As these backgrounds generically involve RR fields, this is 
a highly non-trivial problem.      

One example of this phenomenon is the Chamseddine-Volkov, Maldacena-N\'u\~nez geometry(MN)
\cite{chamsI, chamsII, Maldacena:2000yy}. The field theory is constructed by wrapping $N$ D5 branes on 
the non-trivial two-cycle of the resolved conifold in such a 
way to preserve ${\cal N}=1$ supersymmetry in the resulting 
4D Minkowski space. Upon reduction on $S^2$, the theory in 
4D becomes $SU(N)$ \sym theory with its massless vector multiplet 
coupled to an infinite tower of massive vector and 
massive chiral multiplets
\footnote{A recent study of the classical spectrum is in \cite{Andrews:2005cv}.}.    
The dual 10 D geometry involves a non-trivial three-cycle inside the CY 3-fold 
transverse to the Minkowski space. In this dual picture, the masses of the KK 
modes are inversely proportional to the volume of $S^3$. One can try to decouple 
them in the IR 
by shrinking this volume by adjusting the value of the dilaton at the origin. 
However, in this regime the radius of curvature becomes small and the supergravity 
approximation fails. As a result, the energy scale of the KK-dynamics and the SYM dynamics 
(\eg typical mass of a glueball) is at the same order.       
 
%The geometry that is dual to this field theory is \cite{Maldacena:2000yy} the product of 
%$M^4$ and the resolved conifold that is further deformed by $N$ units of 
%three-form flux on the non-trivial three-cycle of the resolved conifold.
%The mixing problem in this example arises as follows. The curvature radius 
%of the geometry in the IR region (close to the origin) 
%is inversely proportional to $m_0 e^{-\phi_0}$ where $m_0=\sqrt{\a'g_sN}$ is roughly 
%the average glueball mass in the dual theory. On the other hand, the mass of the 
%KK modes are inversely proportional to the volume of the three-cycle which is again 
%$m_0 e^{-\phi_0}$. Therefore the supergravity approximation and the condition of 
%decoupling the KK sector are exclusive. 

A new conjecture on the gauge-gravity correspondence 
was put forward by the recent work of Lunin and Maldacena \cite{Lunin:2005jy}.  
The authors employed an $SL(3,R)$ transformation in order to generate new 
IIB backgrounds from an original one that involves a torus isometry. 
The transformation changes the volume of the torus and introduces other 
changes on the various other fields in the background. According to their 
conjecture, the transformation introduces
certain phases or non-local deformations in the Lagrangian of
the original dual field theory. Employing this idea, they discovered the 
gravity dual of the $\b$ deformed ${\cal N}=4$ SYM\cite{Leigh:1995ep}. 
Further literature on this duality can be found in \cite{bisuru1} and some
applications of the Lunin-Maldacena conjecture to other backgrounds is in 
\cite{Gursoy:2005cn} and \cite{bisuru2}. 

Relevance of the Lunin-Maldacena conjecture to the aforementioned 
KK-mixing problem was pointed out in \cite{Gursoy:2005cn}, 
where a non-singular one-parameter 
subgroup of $SL(3,R)$ with a real parameter $\g$ was applied to the 
MN background. This results in a new background of IIB that involves a 
complicated metric, a non-trivial dilaton, NS form and various RR-forms. 
Henceforth we shall denote the background obtained in \cite{Gursoy:2005cn} as 
\mng. The details of this background are presented in section 2 and Appendix A. 
According to the prescription of Lunin and Maldacena, the deformation 
only affects the particular fields on the $D5$'s 
which are charged under the torus isometry: These fields 
acquire dipole moments that are proportional to $\g$ \cite{Bergman:2000cw}. 
It is observed in \cite{Gursoy:2005cn}
that these charged fields become precisely the KK vector and hyper multiplets upon 
reduction on $S^2$, and that, one obtains 4D \sym that is  
coupled to a KK sector whose interactions among themselves are 
altered by the deformation parameter $\g$. It is a local field theory in 4D. 
Most importantly, the pure SYM sector is left intact under the deformation.  
Various checks of this idea have been made in \cite{Gursoy:2005cn}. 
In particular, gravity calculations show that
the new theory confines with the same string tension, 
and that the NSVZ $\b$-function and $\Theta_{YM}$ of \sym are the same 
as obtained from the MN background \cite{DiVecchia:2002ks}. Thus, the quantities 
that are inherent to \sym are independent of $\g$. 
On the other hand, the quantities that are inherent to the KK-sector, such 
as the masses of the KK modes are shown to transform under the deformation.

The general conclusions of \cite{Gursoy:2005cn} are as follows:
\begin{itemize}
\item The \mng background provides a continuous set of new UV completions 
of the dual field theory that flows to \sym in the far IR, 
\item The $\g$ deformation increases the masses of the KK modes, hence 
help decoupling them from the pure SYM sector
\footnote{It is unclear how useful this is in practice, because in the 
supergravity approximation the parameter $\g$ is restricted to obey $\g R^2<<1$ where 
$R>>1$ is the radius of curvature.}, and more importantly,
\item The parameter $\g$ serves as a marker 
that marks the KK-dependent ``non-universal'' features of the
dual field theory, therefore helps identifying in the gravity description, 
the universal features 
of \sym theory in the IR.
\end{itemize}         
The effects of the dipole deformations on the KK-modes from 
a field theory viewpoint 
is very recently discussed in \cite{Landsteiner:2006qb}. 
Further literature that focused on the 
relevance of the Lunin-Maldacena conjecture for the KK-mixing problem   
can be found in \cite{bisuru3}.  

In this paper, we further investigate the third highlighted 
feature above. We provide an illustration of this idea by constructing new quantities from the 
\mng background that depend and do not depend on $\g$, as follows. 
We consider the Penrose limit 
of the background that focuses on a null geodesic in the IR. This results in a pp-wave, 
that is the geometry seen by strings moving very fast along the equator 
of the $S^3$ at the origin. This limit was previously studied by Gimon
et al. in \cite{Gimon:2002nr} in case of the original MN background. They have found an 
exactly solvable pp-wave whose light-cone Hamiltonian is conjectured to be dual to 
the non-relativistic Hamiltonian that describes the dynamics of what the authors 
called the ``annulons''. 
These objects are long and massive gauge-invariant operators built out of the 
KK-fields in the geometry and they carry large $U(1)$ charge $J$ (dual to the angular 
momentum along the geodesic). The 8 transverse directions of the 
MN pp-wave\cite{Gimon:2002nr} are the three flat directions that correspond to the 3 spatial
directions of $M^4$ where the KK-hadrons can move, two massive directions of mass
$m_0/3$ and two other massive directions of mass $m_0$ where $m_0$ 
is the typical glueball mass. There is also another ``accidentally'' 
massless direction that is believed to be an artifact of the Penrose limit and has no 
relevance for the dual field theory physics. Further literature on the ``annulons'' is 
in refs.\cite{bisuru4}. 

We review the MN pp-wave geometry in section 3.1 and obtain the generalization
of their construction to the \mng geometry in the sections 3.2-3.4. 
This generalization proved technically hard both because the 10D \mng metric 
is much more complicated than the original MN metric and because there are more 
fields present in the background. We show that, in order to obtain a non-trivial pp-wave
one should also rescale the deformation parameter as (this is similar to the pp-wave 
constructed in \cite{Lunin:2005jy}),
\be\lab{rescale}
\g\to 0, \quad J\to\infty, \quad \g J=\gt=const.
\ee 
We work out the 
Penrose limit of the metric in section 3.2 and obtain the limits of the NS form 
and the various RR-forms 
in sections 3.3 and 3.4. 
In particular, we show that the deformation parameter in the pp-wave limit 
can alternatively be thought as the value of the axion in the geometry: 
\be\lab{ax}
\chi = \frac{\g}{4}.
\ee 

As the light-cone Hamiltonian of this pp-wave is conjectured
to describe the dynamics of the hadrons that are built out of the KK-modes 
in the dual field theory \cite{Gimon:2002nr}, one expects to see that the geometry 
depends non-trivially on the dipole moment $\g$. We show that this is indeed the case: 
The masses of the massive directions shift by,
\be\lab{massshift}
M^2\to M^2 + \frac{\gt^2}{4},
\ee
with respect to the pp-wave of the original MN geometry. 

The pp-wave is quadratic, hence exactly solvable. In the sections 4.1 and 4.2 respectively, 
we compute the spectrum of bosonic and fermionic string excitations 
in the light-cone gauge and find that, generically, the spectrum looks
as follows,
\be\lab{spec}
w_n = \sqrt{(n+a\gt)^2 + b M^2},
\ee
where $a$ and $b$ are coefficients that vary in different sectors of the spectrum.
We note that (\ref{spec}) is very similar to the pp-wave spectrum found in \cite{Niarchos:2002fc} and 
\cite{Lunin:2005jy}.         
   
We continue our investigation of the \mng 
pp-wave, by studying the thermal properties of strings in this background. The thermal partition 
function of single-string excitations in the maximally-symmetric pp-wave 
background \cite{Blau:2001ne} has been studied in \cite{PandoZayas:2002hh}. 
It was shown that the partition function 
becomes ill-defined at a Hagedorn temperature which varies according to the mass parameter.
Similarly, the UV asymptotics of the MN pp-wave of \cite{Gimon:2002nr} 
was studied in \cite{PandoZayas:2003jr}. 

In section 6, we compute the thermal partition function of single-strings in our geometry 
and in section 7 we derive the Hagedorn temperature. Quite surprisingly, this temperature
turns out to be independent of the deformation parameter, although the partition 
function exhibits non-trivial dependence on $\gt$. Thus, 
according to our general discussion above, the Hagedorn temperature may serve as a 
non-trivial ``universal'' quantity, that is relevant for the pure SYM dynamics. 
A discussion of this and further points follow in the last section. 
Three appendices detail our computations. 
       
\section{The \mng Background}
\setcounter{equation}{0}

The background of \cite{Gursoy:2005cn} 
is conjectured to be dual to ${\cal N}=1$ SYM with 
a new infinite family of UV completions parametrized by 
$\g\in {\mathbf R}$. The metric of \cite{Gursoy:2005cn} can be put 
in the following simpler and useful form (in the string frame):
\be\lab{met}
ds^2 = ds^2_{0} + ds^2_{\g}.
\ee   
The first term is the original Chamseddine-Volkov metric\cite{chamsI, chamsII,
  Maldacena:2000yy},
\be\lab{mn}
ds^2_{0}\,=\,e^{\phi}\,\,\Big\{\, 
dx^2_{1,3}\,+\,\frac{1}{m_0^2}\left(
dr^2\,+\,e^{2h}\,\big(\,d\theta^2+\sin^2\theta 
d\varphi^2\,\big)\,+\,{1\over 4}\,(w^i-A^i)^2\right)\,\Big\}\,\,,
\ee
where $\phi$ is the dilaton in the MN background (App. A). 
The parameter that appears in the metric is 
the typical glueball mass, $m_0^2 = (\a' g_s N)^{-1}$. 
The gauge fields $A^i$ determines the fibration of the $S^2$ on the $S^3$
and $w^i$ are the left invariant $SU(2)$ one-forms (for details, see App. A).   

The second part of (\ref{met}) is a five dimensional metric on the angular variables
in the geometry:
\be\lab{metg}
ds^2_{\g} = \!\!\!\left(\frac{e^{\phi}}{4m_0^2}\right)^3\!\!\! 
\frac{\g^2}{1+ \g^2 F^2}\,\, d\W_5^2(r,\q,\qt,\j).
\ee
The explicit forms of $F$ and $d\W_5$ are presented in App. A. 

Note that this form of the metric in (\ref{met}) is useful because 
it clearly isolates the gamma-dependent part in the geometry. It also involves 
many simplifications with respect to the original form given in \cite{Gursoy:2005cn}.
The parameter $\g$ in the metric only appears as in (\ref{metg}).

Apart from the metric, there is a non-trivial dilaton, axion and various forms 
in the geometry\cite{Gursoy:2005cn}. The dilaton and the axion are given by,
\be\lab{dilg}
e^{\phi_\g} = \frac{e^{\phi}}{\sqrt{1+\g^2F^2}},
\qquad\chi = -\frac{\g}{4m_0^2} g ,     
\ee
where the function $g(r,\q,\qt,\j)$ is defined in (\ref{metf}).    

The original MN solution is supported by a 3-form RR flux $F^{(3)} = d C_0^{(2)}$ 
on the non-trivial $S^3$ in the geometry. An expression for $C_0^{(2)}$ is given in 
the appendix A. After the $\g$ deformation it gets deformed into,
\be\lab{RR2g}
C^{(2)}_{\g} = C^{(2)}_0 - \frac{1}{4m_0^2} \frac{\g^2\,F^2}{1+\g^2F^2} \, g\, 
(d\f + \ca_1)\wedge(d\ft + \ca_2).
\ee
The first term is the original MN flux and 
we define the one-forms, $\ca_1$ and $\ca_2$ in appendix A.
Again, this form of the two-form involves many simplifications over 
the original expression given in \cite{Gursoy:2005cn} 
\footnote{There is a mistake in the original expression of $C^{(2)}$ in \cite{Gursoy:2005cn}. 
We thank C. Ahn for pointing this out.} 
It also has the advantage of clearly displaying the $\g$ dependent piece. 

There is also a four-form flux which by using the expressions 
in the Appendix A of \cite{Gursoy:2005cn} can be put in the following form, 
\be\lab{RR4g}
C^{(4)}_{\gamma} = -C^{(2)}_{\gamma}\wedge B^{(2)}_{\gamma} = -C^{(2)}\wedge B^{(2)}_{\gamma}. 
\ee  
where $B^{(2)}_{\gamma}$ is the NS-NS two-form flux that is absent in the original MN 
background but generated after the $\g$ transformation in the deformed geometry. 
It is given as,
\be\lab{NS2g}
B^{(2)}_{\gamma} = \frac {\g F^2}{1+\g^2F^2} (d\f + \ca_1)\wedge(d\ft + \ca_2).
\ee    

Although the background is quite complicated, 
we show in the following section that it admits a nice Penrose limit which results in a simple, 
quadratic pp-wave.  

\section{The Penrose Limit}
\setcounter{equation}{0}

\subsection{The pp-wave of the original MN background}
 
It was argued in \cite{Gimon:2002nr} that the dynamics of the KK-hadrons in the gauge theory is 
captured by the pp-wave geometry that is obtained from the original metric by 
focusing on a geodesic on the $S^3$ at IR. Recall that the original MN metric has a 
non-trivial $S^3$ at the origin and this fact is unchanged by the $\g$ deformation, as can 
be seen from (\ref{met}) and (\ref{metg}). 
In this section we first recall the Penrose limit of the original MN geometry \cite{Gimon:2002nr}
and then we apply the same steps to the more complicated ${\rm MN}_{\g}$ geometry. 
Our limit focuses on the same geodesic as in \cite{Gimon:2002nr}, hence we shall be able to compare 
the resulting pp-waves in the original and the deformed MN geometries and learn about the precise 
effects of the deformation on the field theory dynamics. 

The Penrose limit of an arbitrary Lorentzian space-time \cite{Penrose} 
as generalized to supergravity by \cite{Gueven:2000ru} utilizes the 
homogeneity property of the supergravity action under 
constant scalings of the metric and the various other fields in the geometry. 
Therefore, the best way to keep track of the scalings of the fields in our geometry 
is to look at the IIB SG Lagrangian:
\be\lab{SG}
{\cal L} = \frac{1}{2\k_{10}^2}e^{-2\phi}\left(-{\cal R}*1+4d\phi\wedge *d\phi-\half H_3\wedge * H_3\right) 
-  \frac{1}{4\k_{10}^2}\left(d\chi \wedge * d\chi+\tilde{F}_3 \wedge * \tilde{F}_3
+ \half \tilde{F}_5 \wedge * \tilde{F}_5\right). 
\ee
The combinations of the forms that appear in (\ref{SG}) are defined as,                
\be\lab{RRdef}
\tilde{F}_3 = dC_2 -\chi H,\,\,\,\, \tilde{F}_5 = dC_4 -\half C_2\wedge H 
+\half B_2\wedge F_3 + \half B\wedge B\wedge d\chi.  
\ee
Let us now recall the pp-wave limit of the first part in (\ref{met}), \cite{Gimon:2002nr}.  
The authors of \cite{Gimon:2002nr} introduce a coordinate patch near a geodesic 
on $S_3$ at the origin by the following change of variables in (\ref{mn}):
\be\lab{pp}
\vec{x} \to \frac{\vec{x}}{R}, \,\, r\to \frac{m_0}{R} r, \,\, \qt \to \frac{2m_0}{R} v. 
\ee
The limit, then is defined by setting the value of the dilaton at origin as 
\be\lab{dilorg}
e^{\phi_0} = R^2,
\ee
and taking $R\to\infty$ in (\ref{pp}) and (\ref{dilorg}). However, the computation is not
straightforward because the one-forms $A_i$ in (\ref{mn}) blow-up as $R\to\infty$. This is because 
the gauge field $A^i$ is pure gauge at the origin:
\be\lab{pg}
A = -i dh\, h^{-1} + {\cal O}(r^2), \,\, h= e^{-i \s^1 \frac{\q}{2}}e^{-i \s^3 \frac{\f}{2}}.
\ee  
Gimon et al. solve this problem by gauging away the pure gauge part of $A$ by the following
change of variables on $S^3$\footnote{In \cite{Gimon:2002nr}, 
this was done by a gauge transformation on $A$ 
in the 7D picture. However we need the picture where we view this transformation 
as a change of variables in the 10D.}: 
\be\lab{cov}
g\to h\,g, \quad g = e^{-i \s^3 \frac{\j}{2}}
e^{-i \s^1 \frac{\qt}{2}}e^{-i \s^3 \frac{\ft}{2}}.  
\ee

After this is done, the new gauge field $A'$ becomes ${\cal O}(r^2)$. 
Therefore the limit 
that scales $r\to 0$ in (\ref{mn}) becomes well-defined. 
This trick amounts to finding a better 
parametrization of the 10D solution that is suitable for the pp-wave limit in question. 
Calling the new angular variables the same as the original ones, Gimon et al. 
then introduce a shift in the angles to 
absorb various angular cross terms in the resulting metric (see \cite{Gimon:2002nr} for details): 
\be\lab{shift}
\f \to \f - \frac16 (\j + \ft), \,\, \ft\to \ft + \half(\j + \ft).
\ee
Finally, one defines the light-cone coordinates,
\be\lab{lcc}
x^+ = x^0, \,\, x^- = \frac{R^2}{2}(x^0 - \frac{\j + \ft}{2m_0}),      
\ee
and takes the Penrose limit, $R\to\infty$. 
This results in the following linear, quantizable pp-wave geometry\cite{Gimon:2002nr}:
\be\lab{pp0}
ds^2_0 = -4dx^+dx^- +d\vec{x}^2 + d\vec{y}^2 + d\vec{z}^2 
- m_0^2(dx^+)^2\left(z_1^2+ z_2^2 + \frac19 y_2^2 +\frac19 y_3^2\right).
\ee
Here $\vec{z} = (z_1,z_2)$ and $\vec{y} = (y_1,y_2,y_3)$ are coordinates on 
${\mathbf R}^2$ and ${\mathbf R}^3$. They are related to the original variables 
in the MN geometry as follows:
\bea
y_1 &=& r \cos{\q},\,\, y_2 = r \sin{\q}\cos{\f},\,\, y_3= r \sin{\q}\sin{\f},\nn\\  
z_1 &=& v\cos{\ft},\,\, z_2 = v\sin{\ft}.
\eea 
We see that there are three massless coordinates that 
correspond to the spatial coordinates of the original Minkowski space, 
two massive coordinates with masses $m_0^2$, two other massive coordinates 
with mass $\frac19 m_0^2$ and one ``accidentally'' massless coordinate, $y_3$. 
We have more to say about this massless coordinate below. 

Having established the steps that lead to a well-defined pp-wave metric above, one can 
easily work out the limit of the 3-form in the MN geometry\cite{Gimon:2002nr}:
\be\lab{MNc2pp}
C^{(2)}_{pp} = \frac{2m_0}{R^2} x^+\Big\{dz_1\wedge dz_2
+\frac13 dy_2\wedge dy_3\Big\}.
\ee  

We note that the scaling of various objects in the geometry with $R$ is consistent 
with (\ref{SG}): The metric (\ref{mn}) scales as $R^0$, using (\ref{dilorg}) and (\ref{pp}).
Thus ${\cal R}\sim R^0$ and the first part of (\ref{SG}) scales as $R^{-4}$. Similarly, 
the second part of (\ref{SG}) scales as $R^{-4}$, using (\ref{MNc2pp}). 
Therefore the overall factor of $R^{-4}$ in (\ref{SG}) can consistently be absorbed as 
\be\lab{kredef}
\k^2_{10}\to\k^2_{10} R^4.
\ee     

\subsection{Penrose limit of the metric in \mng}

Now, we apply the same steps to the second part of the metric in 
(\ref{met}). At first sight, it seems hopeless to obtain 
a linear pp-wave geometry in the same limit, (\ref{pp}) as $ds^2_{\g}$ 
is extremely non-linear. However we will show below that this first 
impression is wrong. 

One first carries out the change of variables (\ref{cov}) 
in the second part of (\ref{met}). One needs the explicit form of the 
new angles in terms of the old ones. Let us denote the new variables
with bars upon them.   
We present the expressions that relate new angles $\oq$, $\oj$ and $\of$ to 
the old ones, $\qt$, $\j$ and $\ft$ in Appendix B. 

In practice, one only needs the expressions for the leading order terms in the expansions 
of $\sin{\oq}$, $\cos{\oq}$, $\sin{\oj}$ and $\cos{\oj}$ functions 
that appear in (\ref{metf}) in the variable $\qt$. This is
because we will eventually take the Penrose limit $R\to\infty$ in (\ref{pp}). 
We present these expansions also in the Appendix B. 

Inserting these expansions given by (\ref{covs}) in (\ref{metf}) 
one obtains a nice result after many simplifications: 
$d\W_5^2$ in (\ref{metg}) becomes order $1/R^2$ ! 
In detail, after the redefinitions in (\ref{pp}), this angular piece 
becomes,
\be\lab{dw5pp}
d\W_5\to 
\frac{4m_0^2}{R^2} v^2\sin^2(\j-\f)(d\q-d\oq)^2 
+\frac{4m_0^2}{R^2}\left(v^2+r^2\sin^2{\q}\right)(d\f+d\of)^2 + {\cal O}
\left(\frac{1}{R^2}\right)d\oj(d\oj+\cdots).
\ee 
From this observation, it follows that one can define 
a non-singular pp-wave limit (\ref{pp}) also in (\ref{metg}) if we 
also rescale $\g$ such that\footnote{One can also try $\g R^3 = fixed$ but this 
gives exactly the original PP-wave of MN as the $\g$ dependence vanishes.}, 
\be\lab{rg}
\gamma\to 0,\,\, R\to\infty, \,\, \gt = \g R^2 = fixed.
\ee
We also note that $f-g^2 \sim {\cal O} (1/R^2)$, hence the denominator of the 
term in front of $d\W_5^2$ in (\ref{metg}) becomes 1. 

When one applies the change of variables in (\ref{covs}) in the metric, 
one should also apply them to the one-forms $d\j$, $d\ft$ and $ d\qt$. It should 
now be clear that we only have to keep the zeroth order terms:
\be\lab{dcovs}
d\oq = d\q +{\cal O}(1/R), \,\, d\of = d\ft + d\j - d\f +{\cal O}(1/R), 
\,\, d\oj = {\cal O}(1/R).
\ee  
In particular all the terms that multiply $d\oj$ in (\ref{dw5pp}) go away
in the limit. Another nice surprise that follows from (\ref{dcovs}) is that 
the first term in (\ref{dw5pp}) disappears! 
If this did not happen we would get a non-linear term $\sin^2(\j-\f)$. 

Having carried out the change of variables, we call the new angles 
$\oj$, $\oq$ and $\of$ back as $\j$, $\qt$ and $\ft$.
The next steps are to apply the same shift in the angles as in (\ref{shift})
define the light-cone coordinates in (\ref{lcc}) and take the limit $R\to\infty$   
in $ds^2_{\g}$. These steps are simple and one arrives at the following 
contribution from the second part of the metric in (\ref{met}):
\be\lab{ppg}
ds^2_{\g} \to -\frac{\gt^2}{4m_0^2} (dx^+)^2 
\left(z_1^2 + z_2^2 + y_2^2 + y_3^2\right). 
\ee  
We combine (\ref{pp0}) with (\ref{ppg}) and obtain our final expression:
\be\lab{ppf}       
 ds_{pp}^2 = -4dx^+dx^- +d\vec{x}^2 + d\vec{y}^2 + d\vec{z}^2 
- m_0^2(dx^+)^2\left((1+\frac{\gb^2}{4})(z_1^2+ z_2^2) 
+ (\frac19 + \frac{\gb^2}{4})(y_2^2 + y_3^2)\right).
\ee
Here we introduced the dimensionless quantity $\bar{\gamma}$:
\be\lab{barg}
\gb = \frac{\gt}{m_0^2}.
\ee 

We observe that the masses of the massive directions are shifted 
by terms ${\cal O} (\gt^2)$ under the transformation. This is in 
accordance with our expectations: 
According to the conjecture of \cite{Gimon:2002nr}, the
 pp-wave essentially carries information about the hadrons made out of the KK-modes. 
If this conjecture is correct, we expect that the massive directions get deformed 
in the pp-wave of the \mng geometry. 
We indeed observe this in (\ref{ppf}). Actually, one has to be more careful in 
identifying the precise effect of the $\g$ deformation\footnote{We are grateful to Elias Kiritsis 
for taking our attention to this point.}. The mass shift in one 
direction can be absorbed by a constant rescaling of the coordinates. It is the 
change in the ratio of two massive directions, that has an invariant meaning. 
We see in (\ref{ppf}) 
that the ratio of the $y_{2}$, $y_{3}$ to the $z_{1}$, $z_{2}$ directions indeed 
change under this process as, 
\be\lab{chrat}
\frac{m_y^2}{m_z^2} = \frac19\to\frac{1/9+\gb^2/4}{1+\gb^2/4}.
\ee              
Moreover, the 
massless directions are protected. One indeed expects no change in the 
three massless directions $x_i$ as they are protected 
by space translation symmetries in the original MN model 
and also in the $\g$ deformed theory. However the masslessness of $y_1$ 
not being lifted by the $\g$-deformation is disturbing. This implies 
that the ``accidental'' symmetry that protected $y_1$ is still there 
in the Penrose limit of the $\g$-deformed theory. 
It is believed that this accidental symmetry of the MN hadrons 
is just a non-universal feature of the model \cite{Gimon:2002nr}. 
We also believe that the same feature in the 
$\g$-deformed theory is also non-universal and 
is not expected to arise for example neither 
in the pp-wave limit of the $\g$-deformed Klebanov-Strassler (KS) theory 
\cite{Klebanov:2000hb}, \cite{Gimon:2002nr, Lunin:2005jy} 
or in the $\g$-deformed G2 model \cite{Bertoldi:2004rn}. 

However, a more striking observation follows from (\ref{ppf}): 
Comparison of the pp-waves of MN and KS models led the authors of \cite{Gimon:2002nr} 
identify the two massive directions $z_1$ and $z_2$ 
---which happened to have the same mass, $m_0^2$ both in the MN and the KS model--- 
as exhibiting a common feature that is expected to be inherent to the ``hadron'' 
sector in the gravity duals of ${\cal N}=1$ SYM. These directions were called ``universal'' 
in \cite{Gimon:2002nr} and in some further literature. Here, 
we see that this statement should be modified. 
The $\g$-deformed model also flows to ${\cal N}=1$ SYM in the far infrared, 
however the masses of $z_1$ and $z_2$ here depend on $\g$, 
hence model dependent and 
non-universal. Non-universality of these directions also follow from comparison 
of MN and KS pp-waves to that of the G2 model \cite{Bertoldi:2004rn}. 

\subsection{ Penrose limit of the NS form } 

The construction of the bosonic world-sheet
action of the pp-wave involves the Penrose limit of the NS-form. 
As we show in section 4.1, the NS form contributes non-trivially 
to the frequencies of the higher string modes, $n>0$.

As the computation for the metric is given in detail above, 
we shall only present the final result for the NS form here:
\be\lab{B2pp}
B^{(2)}_{pp} = \frac{\gb m_0}{2}
\left(y_2 d y_3 - y_3 d y_2 + z_2 dz_1 - z_1 dz_2\right)\wedge dx^+.
\ee     
The dimensionless parameter $\gb$ is defined in (\ref{barg}).  
This is given in a gauge convenient for construction of the world-sheet action. 
The gauge independent object is the NS-3 form flux:
\be\lab{H3pp}    
H^{(3)}_{pp}  = \gb m_0 (d y_2\wedge d y_3 -d z_1\wedge d z_2)\wedge dx^+.
\ee
\subsection{ Penrose limits of the RR forms }\lab{secRR}
 
In the Penrose limit (\ref{rg}), the axion becomes,
\be\lab{axpp}
\chi_{pp} = \frac{\gb}{4}\frac{1}{R^2}.
\ee
Therefore {\em one can alternatively think of the deformation parameter $\g$ as 
the expectation value of the axion in the PP-wave geometry.} 
We note that in the Lagrangian (\ref{SG}) and in the equations of motion, only 
the combination $\chi_{pp}R^2$ appears which stays finite in the limit. The same 
applies to the RR three-form below. 

The limit of the RR two-form (\ref{RR2g}) is very easy to obtain. The pp-limit 
of the first term in (\ref{RR2g}) was already obtained in \cite{Gimon:2002nr}. On 
the other hand the second part is proportional to the NS-form (\ref{NS2g}) 
thus we can directly use our result in the previous subsection. The result is, 
\be\lab{C2pp}
C^{(2)}_{pp} = \frac{2m_0}{R^2} x^+\Big\{(1-\frac{\gb^2}{8})dz_1\wedge dz_2
+(\frac13+\frac{\gb^2}{8})dy_2\wedge dy_3\Big\}
\ee 
or equivalently, 
\be\lab{F3pp}
F^{(3)}_{pp} = \frac{2m_0}{R^2} dx^+\wedge\Big\{(1-\frac{\gb^2}{8})dz_1\wedge dz_2
+(\frac13+\frac{\gb^2}{8})dy_2\wedge dy_3\Big\}.
\ee 

The RR four-form is given by (\ref{RR4g}). In the limit this becomes,
\be\lab{C4pp1} 
C^{(4)}_{pp} = -C^{(2)}_{pp}\wedge B^{(2)}_{pp} = 0,
\ee
where we used (\ref{B2pp}) and (\ref{C2pp}). The fact that the four-form vanishes 
in the pp-wave limit is quite a nice simplification for the spectrum of string oscillators
in this geometry. In particular, the masses of the fermionic oscillators do not acquire 
contribution from the 5-form. The same is true for the pp-wave 
\cite{Gimon:2002nr} that follows from the Klebanov-Strassler (KS) 
geometry \cite{Klebanov:2000hb}. 

The combinations of forms that enter the equations of motion and the supersymmetry variation 
equations in IIB are $\tilde{F}^{(3)} = F^{(3)} - \chi H^{(3)}$ and 
$\tilde{F}^{(5)} = d C^{(4)} -\half C^{(2)}\wedge H^{(3)} - \half B^{(2)}\wedge F^{(3)} 
-\half B\wedge B\wedge d\chi$. 
Using the above expressions we obtain, 
\be\lab{F3tpp}
 \tilde{F}^{(3)} =  \frac{2m_0}{R^2} dx^+\wedge(dz_1\wedge dz_2+\frac13 dy_2\wedge dy_3),
\ee 
and 
\be\lab{F5tpp}
 \tilde{F}^{(5)} = 0.
\ee 
Interestingly, $\tilde{F}^{(3)}$ has exactly the same expression as the $\tilde{F}^{(3)}$
of the pp-wave of the original MN geometry \cite{Gimon:2002nr}: {\em the gamma 
deformation does not affect the three-form flux} $\tilde{F}^{(3)}$. 
This fact also has important implications for the fermionic string spectrum. 
We also note that the combination $\tilde{F}_3$ that appears in (\ref{SG}) comes out with a 
homogeneous scaling with $R$. This fact makes the Penrose limit well-defined and also provides
a check on our computations. One again checks that both the first and the second parts of (\ref{SG}) 
scale as $R^{-4}$, hence this factor can be absorbed into the redefinition (\ref{kredef}).    

In summary, the new pp-wave geometry is given by eqs. (\ref{ppf}), (\ref{axpp}), 
(\ref{H3pp}),(\ref{F3pp}) and  (\ref{F5tpp}). 
This is a result of a long and tedious computation, hence it is crucial to confirm 
our findings by checking the equations of motion. The only non-trivial Einstein equation 
in this pp-wave geometry is, 
\be\lab{eompp}
R_{++} = \frac{1}{4}(H_{+ij}H_+^{ij}-\frac{1}{12}H_{ijk}H^{ijk})
+\frac{e^{2\phi}}{4}(\tilde{F}_{+ij}\tilde{F}_+^{ij}-\frac{1}{12}\tilde{F}_{ijk}\tilde{F}^{ijk}).
\ee
It is straightforward to verify that both the RHS and LHS of this equation equal 
the sum of the masses, $\sum_i m_i^2 = m_0^2(\frac{20}{9} + \gb^2)$. 

\section{ The Spectrum} 
\setcounter{equation}{0}

\subsection{The Boson spectrum}

In this section we derive the bosonic part of the world-sheet 
action in the light-cone gauge and compute the spectrum of the 
bosonic string modes. The bosonic action is, 
\be\lab{wsa}
S = \int d^2 \s \sqrt{g} \left(g^{ab} \6_aX^{\m} \6_b X^{\n} G_{\m\n} 
+\eps^{ab} \6_a X^{\m}\6_b X^{\n}B_{\m\n}\right), 
\ee
where $G_{\m\n}$ and $B_{\m\n}$ are given by equations (\ref{ppf}) and (\ref{B2pp}).    

The symmetries of the action are fixed by the light-cone gauge condition:
\be\lab{LCG}
g_{ab} = diag(-1,1),\qquad X^+ = \a'p^+ \t.
\ee
Inserting this in (\ref{wsa}) immediately shows that we have 4 massless oscillators, 
$x_1$, $x_2$, $x_3$ and $y_1$ and two systems of coupled oscillators 
$y_2$, $y_3$ and $z_1$, $z_2$. The coupling is due to the non-trivial NS-form. 
The same coupling arises in the pp-wave constructed from the KS solution \cite{Gimon:2002nr}. 
The coupled systems are easily solved by introducing the mode expansions,
\be\lab{exps}
y_2 = \sum_{n} A_n e^{(i w_n \t + n\s)},\qquad y_3 = \sum_{n} B_n e^{(i w_n \t + n\s)},
\ee
and similarly for $z_1$ and $z_2$. One obtains the following spectrum for $y_2$ and $y_3$ 
\be\lab{specy}
(w_n^{y})^2= \frac{\mb^2}{9}+(\frac{\mb\gb}{2}\pm n)^2,
\ee
and for $z_1$ and $z_2$,
\be\lab{specz}
(w_n^z)^2 = \mb^2+(\frac{\mb\gb}{2}\pm n)^2, 
\ee    
where we introduced the following 
dimensionless parameter,
\be\lab{mb}
\mb = m_0 \a' p^+,
\ee
and $\gb$ is defined in terms of $\gt$ as in (\ref{barg}). 
In addition to (\ref{specy}) and (\ref{specz}), 
there are of course the massless directions $x_1$, $x_2$, $x_3$ and $y_1$ 
in (\ref{ppf}) with frequencies 
\be\lab{specx}
w^0_n = n.    
\ee

A number of observations follow from (\ref{specy}) and (\ref{specz}). 
First of all, the gamma-deformed spectrum goes 
over to the original spectrum of the MN pp-wave \cite{Gimon:2002nr} in the limit 
$\gb\to 0$. One also observes that, as in the case of the 
KS pp-wave, the presence of NS two-form does not affect the zero frequencies. 

\subsection{ The Fermion spectrum} 

The fermionic part of the world-sheet action in a general pp-wave background is obtained in
\cite{Metsaev:2001bj}, \cite{Russo:2002rq} (see also \cite{Brecher:2002ar}) from the 
Green-Schwarz action in the light-cone gauge,
\be\lab{lcgf}
\Gamma^+\q^I = 0,
\ee
where $\q^I$, $I=1,2$ are the 16 component Majarona-Weyl spinors of IIB. 
The fermion action becomes,  
\be\lab{alcf}
S_F = \frac{i}{\pi}\int d\s^2\left(\eta^{\a\b}\d_{IJ}-\e^{\a\b}\r^3_{IJ}\right)
\6_{\a}X^{\m}\6_{\b}X^{\n}\bar{\q}^I\Gamma_{\m}{\cal D}_{\n}\q^J.
\ee 
The covariant derivative, ${\cal D}$, which also appears in the supersymmetry variation 
of the gravitinos is (for constant axion and dilaton and vanishing 5-form),
\be\lab{covD}
{\cal D}_{\m} = D_{\m} -\frac{e^{-\phi}}{96} \sH_{\m}\s_3-\frac{1}{96} \tilde{\sF}_{\m}\s_1,
\ee
with 
\be\lab{defs}
\sH_{\mu} = H_{\n\r\s}\Gamma_{\m}^{\n\r\s} - 9 H_{\m\n\r}\Gamma^{\n\r}
\ee
and similarly for $\tilde{F}$. 

As first observed in \cite{Russo:2002rq}, the RR 3-form contributes as a mass term 
for the fermions and the NS 3-form couples the two sets of fermionic oscillators via 
a chiral interaction term. In our linear pp-wave geometry, the action is still quadratic, 
thus it is straightforward to compute the spectrum. One substitutes the expressions for 
the RR forms in section \ref{secRR} in (\ref{alcf}) and derives the equations of motion for
$\q^I$. To solve the system one combines them as, $\e =\q^1 + i \q^2$, and Fourier expands, 
\be\lab{expsf}
\e = \sum_n \e_n(\t)e^{in\s}.
\ee
The equation of motion, then takes the following form: 
\be\lab{eomf}
\ddot{\e}_n  = \left(\tilde{\sF}^2 + (\sH-in)^2+i[\tilde{\sF},\sH]\right)\e_n.
\ee
Here we defined
\be\lab{ssh}
\sH = -\frac18 p^+\a'\Gamma^{ij}H_{+ij},
\ee
and similarly for $\tilde{F}$. We first note from eqs.(\ref{H3pp}) and (\ref{F3tpp}) that 
the commutator term in (\ref{eomf}) drops. This is in contrast to what happens in the 
two other similar examples of the pp-wave geometries with non-trivial 3-forms that appeared 
in the AdS/CFT literature, namely the  pp-wave of the KS geometry \cite{Gimon:2002nr} 
and the pp-wave obtained in \cite{Brecher:2002ar}. As a result, the spectrum is simpler than 
that of those. The eq.(\ref{eomf}) becomes, 
\be\lab{eomff}
\ddot{\e}_n  = \left\{\frac{\mb^2}{4}\left(-\frac{10}{9}-\frac{\gb^2}{2}
+(\frac23-\frac{\gb^2}{2})\Gamma^{1234}\right)
-n^2+\frac{i}{2}n\mb\gb\left(\Gamma^{12}-\Gamma^{34}\right)\right\}\e_n.
\ee   
Here we denote the $z_1$, $z_2$, $y_2$ and $y_3$ directions in (\ref{ppf}) 
as 1,2,3 and 4, respectively. 

Solving (\ref{eomff}) is a standard exercise performed by expanding $\e_n$ in the
eigenbasis of $i\Gamma^{12}$ and $i\Gamma^{34}$:
\be\lab{sbas}
i\Gamma^{12}|\pm\cdot\rangle = \pm|\pm\cdot\rangle,\quad  
i\Gamma^{34}|\cdot\pm\rangle = \pm|\cdot\pm\rangle.
\ee
One obtains the following sets of doubly degenerate frequencies 
(double degeneracy is because $\e$ is complex),
\bea
(w_n^{++})^2 &=& (w_n^{--})^2 = \frac49 \mb^2 + n^2,\lab{ferpp}\\
(w_n^{+-})^2 &=& \frac{\mb^2}{9}+(\frac{\mb\gb}{2}-n)^2,\lab{ferpm}\\
(w_n^{-+})^2 &=& \frac{\mb^2}{9}+(\frac{\mb\gb}{2}+n)^2.\lab{fermp}
\eea
We note that the fermion spectrum reduces to that of the original MN pp-wave 
in \cite{Gimon:2002nr}, when $\gb$ is taken to zero. 
This is an important check on the computation. 
We also see that the spectrum is invariant under $\gamma\to-\gamma$ as it should be. 
Flipping the sign of $\gamma$ can be undone by a trivial redefinition of chiralities and
relabeling of the coordinates $y_2\leftrightarrow y_3$, $z_1\leftrightarrow z_2$.    

We observe that turning on 
the $\g$ parameter alters the whole spectrum of the original theory except the massless 
sector of the boson spectrum in (\ref{specx}) and the 
$(++)$ and $(--)$ subsectors of the fermion modes. As these subsectors are independent 
of $\gb$ they correspond to a ``universal'' part of 
the field theory that is left invariant under the alteration of the UV 
theory by gamma deformation. 

We demonstrate the shift of the spectrum in fig.1. 
We observe that turning on the $\gb$ parameter, partially removes
some degeneracy in the spectrum of the original pp-wave. Also, and more importantly, 
the shift in the frequencies are positive on average! This fact is in accord 
with the main result of \cite{Gursoy:2005cn}, namely the fact that the deformation 
increases the masses of the KK modes hence help decoupling them from the pure gauge dynamics.
Here we can see explicitly how the spectrum is shifted. It is interesting to note that
the change in the spectrum is somewhat complicated. 
For example in the $w_-$ branch of 
fig.1, the lower modes $n<\mb\gb/4$ are shifted up, whereas all the modes 
$n> \mb\gb/4$ are shifted down. In the $w_+$ branch, all of the modes, $n>0$, are shifted up. 
For every mode that shifts down for $n>\mb\gb/4$ in the $w_-$ branch, there is a corresponding  
mode that shifts up in the $w_+$ branch. 
However, for  $n<\mb\gb/4$ all of the modes shift up. Roughly, the average effect
of turning $\g$ on is to shift about $[\mb\gb/4]$ of the modes up by an amount proportional to 
$(\mb\gb)^2$. Thus, in order to remove more of the modes from the spectrum, 
one clearly needs to increase the scale $\gb\mb$. This shifts the minimum 
of $w_-$ curve in fig.1 toward the right.              

\begin{figure}
 \begin{center}
\epsfxsize=14cm\epsfysize=7cm \epsffile{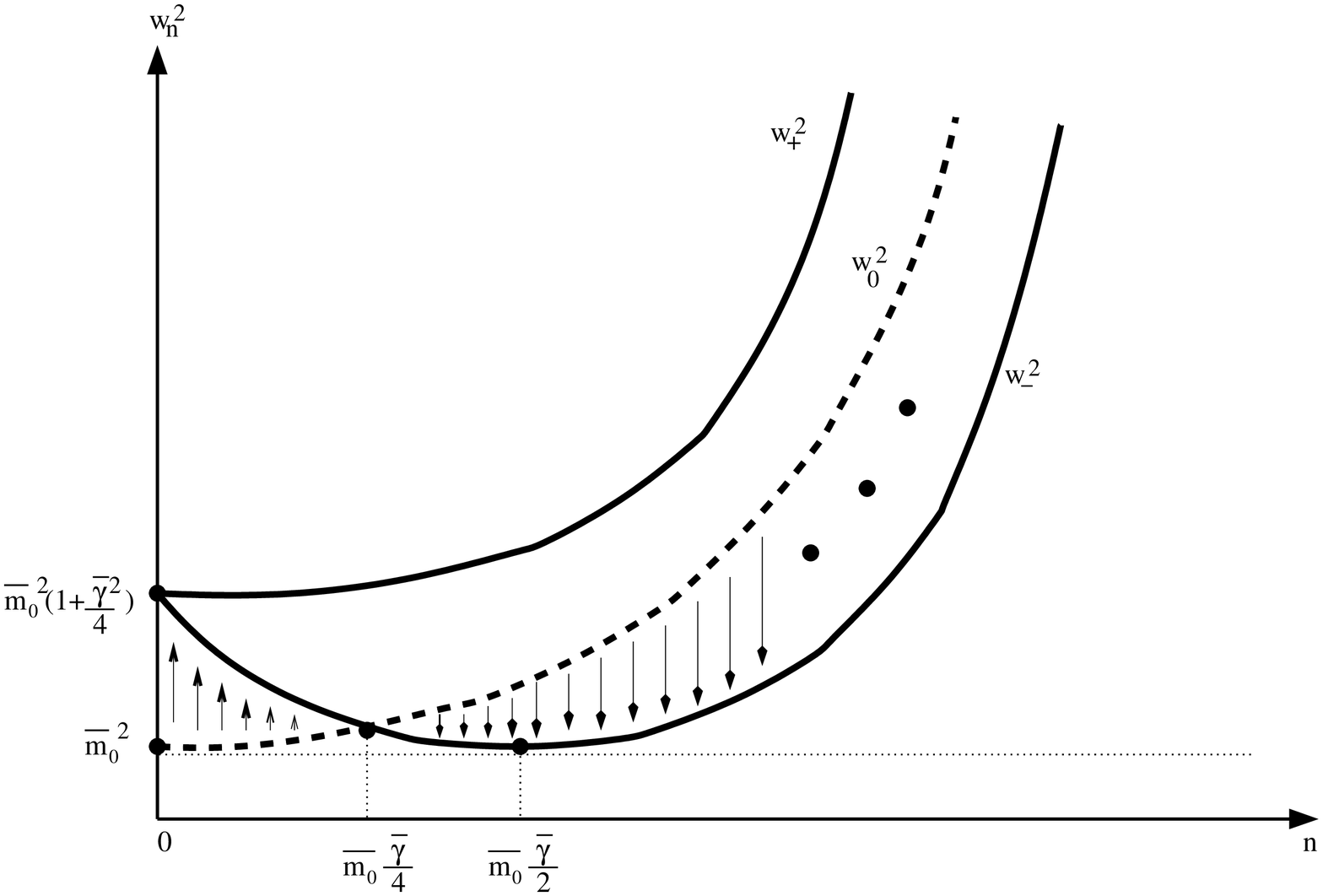}
 \end{center}
 \caption[]{The solid lines show the new spectrum in (\ref{specz}) and the dashed line show the 
old spectrum for $\g=0$. The old spectrum splits into $w_+$ and $w_-$ branches, hence removing the degeneracy partially.
The arrows indicate the shift in some of the modes as $\g$ is turned on. These curves illustrates the shifts 
in the spectra which also takes place in the same way in (\ref{specy}) and in the fermion spectrum.}
 \label{specshift}\end{figure}

In the dual field theory, 
the Penrose limit corresponds to sweeping out all 
of the states in the Hilbert space as $J\to\infty$ except the ones 
that carry finite light-cone momentum and energy\cite{Gimon:2002nr}:
\bea
P^+ &=& \frac{i}{2}\6_-=\frac{m_0J}{R^2} = 
\frac{m_0}{R^2}(J_{\j}+J_{\ft}-\frac 13 J_{\f})= const,\\ 
\lab{lcmom}
H &=& i\6_+ = E - m_0 J  = const. 
\lab{lcen}
\eea
Here, $J_{\a} = -i \6_{\a}$ is the angular momentum in the $\a$ direction.  
In order to derive the relation between the light-cone momentum and the 
angular momentum in (\ref{lcmom}), 
one should recall the relation of the original MN angles 
with the angles used here via the change of variables in (\ref{cov}) and (\ref{shift}).

Let us now understand the supersymmetry properties of the spectrum. 
As clearly seen by comparing the boson and fermion spectra above, 
the ${\cal N}=1$ supersymmetry of the \mng background is 
broken in the pp-wave limit. The reason is as follows \cite{Gimon:2002nr}: 
The charge that is used to take the Penrose limit, namely the $U(1)$ charge $J$, does not 
correspond to an isometry of the original MN background because the only 
$U(1)$ symmetries of the original background are $J_2 = -i\6_{\ft}$ and 
$J_1 = -i(\6_{\f}-\6_{\psi})$ whereas $J = -i\6_{\ft}+i/3\6_{\f}-i\6_{\psi}$.
This is a symmetry that only governs the particular 
sector of the Hilbert space of operators in the dual field theory, 
that is isolated by the rescaling, $J\to \infty$. Therefore, it does not 
commute with the four supercharges of the MN background: supersymmetry is explicitly broken. 
However, $J+2i/3\6_{\f}$ corresponds to a symmetry and we should be able to see the presence of 
two fermionic zero modes in the spectrum if one shifts the Hamiltonian by 
$H\to H +2/3 J_{\f}$, hence the frequencies by $-2\mb/3$. 
We indeed obtain four zero modes from (\ref{ferpp}) by this 
shift, for $n=0$.       

Accordingly, one expects that, although the supersymmetry is broken, the spectrum should 
obey the supersymmetric mass sum rules \cite{Ferrara:1979wa} that requires equality of the 
fermion and boson mass squares in a theory of spontaneously or explicitly broken supersymmetry
(under certain conditions). Indeed, we see that the sum of the 
fermion frequencies in (\ref{ferpp}), (\ref{ferpm}) and (\ref{fermp}) 
and the boson frequencies in (\ref{specy}), (\ref{specz}) and (\ref{specx}) match
order by order in $n$:
\be\lab{sumfreq}
  \sum (w_n^b)^2 =\sum (w_n^f)^2 = \frac{20}{9}\mb^2+n^2+\mb^2\gb^2, \quad \forall n.
\ee
This condition guarantees that the zero-point energy of the corresponding string theory remains 
finite \cite{Bigazzi:2003jk}. In case of the maximally supersymmetric pp-wave, the zero-point energy is not only 
finite but zero. 

Finally, let us discuss an interesting limit of our pp-wave geometry. From the spectrum in 
(\ref{specy}), (\ref{specz}), (\ref{ferpp}), (\ref{ferpm}) and (\ref{fermp}), we see that 
the spectrum further simplifies in the limit,
\be\lab{splim}
     m_0\to 0,\,\, \gt\to 0,\,\, \hat{\gamma} = \frac{\gt}{m_0} = const. 
\ee
This can be achieved by taking $g_s N\to\infty$ 
while keeping $\a'$ and $p^+$ constant  and rescaling $\gt$ as above. We note that 
the spectrum explicitly displays supersymmetry in this limit. Both the boson and fermion 
frequencies come in two types: the flat oscillators $w_n = n$ and the 
``twisted-flat'' oscillators $w_n = |n\pm\a'p^+\hat{\g}|$. From the metric (\ref{ppf}), we
observe that this enhancement in susy is accompanied by 
an enhancement of the transverse isometry to $SO(4)\times SO(4)$. From (\ref{B2pp}), 
(\ref{F3tpp}), we also observe that only the NS two-form is non-trivial. 
It is the simplest type of pp-wave geometry with a non-standard spectrum that involves 
as continuous shift of the flat space frequencies, but without a mass term 
\cite{Russo:1994cv}. In our example, 
we see that the parameter $\g$ of \cite{Russo:1994cv} is 
realized as a dipole deformation in the field theory.

\section{The thermal partition function}
\setcounter{equation}{0}

The canonical ensemble of strings in the light-cone gauge has been studied 
in \cite{Alvarez:1985fw}\cite{Alvarez:1986sj}. 
Investigation of the UV ($p^+\to\infty$) 
behavior of the thermal partition function of string theory in flat space leads   
to the presence of the Hagedorn temperature, $T_H = 1/4\pi \sqrt{\a'}$ 
above of which the theory becomes ill-defined.
 
The thermal properties of string theory in the maximally supersymmetric 
pp-wave background \cite{Blau:2001ne} were analyzed in \cite{PandoZayas:2002hh} and 
\cite{Greene:2002cd} 
where a significantly different behavior is found: The mass parameter in the spectrum 
$w_n = \sqrt{n^2+ m^2}$, depends on $p^+$ as $m = \mu p^+ \a'$, where $\mu$ is the 
value of the five-form in the geometry. This non-trivial dependence on $p^+$ 
alters the UV asymptotics of the thermal partition function and results in a change 
in the Hagedorn temperature \cite{PandoZayas:2002hh}\cite{Greene:2002cd}.  
Depending on the value of $\mu$, the Hagedorn temperature 
ranges from the flat space value $T_H = 1/4\pi \sqrt{\a'}$ for $\mu= 0$ to 
$T_H = \infty$ for $\mu=\infty$. 
The thermal properties is further 
studied in \cite{Brower:2002zx} where it is shown that the free energy of an ideal gas of strings 
in the maximally symmetric pp-wave background stays finite at the Hagedorn temperature, 
hence the Hagedorn behavior possibly signals a phase transition.    

A similar investigation has been carried out for the pp-wave \cite{Gimon:2002nr} 
of the MN background 
in \cite{PandoZayas:2003jr}.
The authors of \cite{PandoZayas:2003jr} showed that the thermal 
behavior of strings in this pp-wave is yet different from the cases above. As the 
glueball mass parameter $m_0$ changes the Hagedorn temperature shifts from 
$T_H = 1/4\pi\sqrt{\a'}$ for $m_0=0$ to $T_H = \sqrt{3}/4\pi\sqrt{\a'}$ 
for $m_0=\infty$ \cite{PandoZayas:2003jr}. Therefore, 
one notes that unlike in case of the maximally symmetric 
pp-wave above, the Hagedorn temperature is bounded as the mass parameter is taken 
to infinity\footnote{See however \cite{Bigazzi:2003jk,Apreda:2003gs}.}. 
In other words, the density of states in this limit behaves 
as 
\be\lab{dos}
\r(E) = \exp(2E/\sqrt{3\a'}). 
\ee
The numerics can be understood as follows \cite{PandoZayas:2003jr}. As the mass parameter 
$m_0$ increases, the gravitational potential in the massive 
transverse directions of the pp-wave becomes steeper. 
As a result, the strings can populate the density of states only by their motion in the 
four flat directions of the MN pp-wave. This explains the relevant factor of 1/2 
in the exponential of (\ref{dos}) with respect to the strings in flat 10D space.       
The paper \cite{PandoZayas:2003jr} also argues that the Hagedorn behavior signals a 
phase transition rather than a thermal limit.  

In this section, we generalize the computations 
in \cite{PandoZayas:2002hh}\cite{Greene:2002cd} and \cite{PandoZayas:2003jr} of the 
thermal partition function for the case of \mng pp-wave. Our motivation is to 
analyze the dependence on the parameter $\g$. This may give us hints about the universal 
thermal properties of the \sym theory in a particular limit that corresponds to the 
Penrose limit. 

We closely follow the notation of \cite{PandoZayas:2002hh}, however we use canonical 
methods to compute the partition function rather than the path integral 
methods in \cite{PandoZayas:2002hh}. The finite temperature single-string partition function
in the light-cone gauge is \cite{Alvarez:1985fw}\cite{PandoZayas:2003jr}, 
\be\lab{pf1}
\cz = \tr\, e^{-\b P^0} = \tr\,e^{-\b (P^+ + P^-)} =
\int_0^{\infty} dp^+\int_{-1/2}^{1/2} d\l\, e^{-\b p^+}\tr\, 
e^{-\frac{\b}{p^+}H + 2\pi i\l(N_L - N_R)}.
\ee 
Here $H$ is the light-cone Hamiltonian and $\l$ is a Lagrange multiplier that 
enforces the level matching condition between 
the left and right moving excitations of the string. 
To avoid cluttering in formulas, below we call the transverse directions 
$y_1$ as 4, $y_2,\, y_3$ as 5, 6 and $z_1,\, z_2$ as 7 and 8. 
The light-cone Hamiltonian is, 
\bea
H &=& \frac{\vec{P}_x^2}{2p^+} + \frac{P_{4}^2}{2p^+} + 
\frac{1}{\a'p^+}\sum_{n=1}^{\infty}w_n^0
\left(\sum_{i=1}^3 N_n^{i} + N_n^{4}\right)
+\frac{1}{\a' p^+}\sum_{n=0}^{\infty}\left(w_n^y(N_n^5+N_n^6)
+w_n^z(N_n^7+N_n^8)\right)\nn\\    
{}&&+ \sum_{n=0}^{\infty}\left(w_n^{++}\sum_{i=1}^4 S_n^i  
+w_n^{+-}(S_n^5+N_n^6)+w_n^{-+}(S_n^7+N_n^8)\right) + H_0\lab{lch}.
\eea
The frequencies are given by the equations (\ref{specx}), 
(\ref{specy}), (\ref{specz}), (\ref{ferpp}), (\ref{ferpm}) 
and (\ref{fermp}). 
$N_n = N_n^l + N_n^r$ is the total number of left and right 
bosonic excitations and $S_n = S_n + \tilde{S}_n$ is the number
of the chiral and anti-chiral fermionic excitations with mode number $n$.   
$H_0$ is the total zero-point energy in the system:
\be\lab{casimir}
 H_0 = 4\Delta^b[0,0] + 2\Delta^b[\mb,\frac{\mb\gb}{2}] +    
2\Delta^b[\frac{\mb}{3},\frac{\mb\gb}{2}] +   
4\Delta^f[\frac{2\mb}{3},0]+
4\Delta^f[\frac{\mb}{3},\frac{\mb\gb}{2}],
\ee
with $\Delta^b[m,\a]$ defined as the Casimir energy of a 
massive complex boson of mass $m$ that satisfies a
twisted boundary condition on the cylinder \cite{Takayanagi:2002pi},
$\phi(\t,\s+\pi) = e^{2\pi i \a}\phi(\t,\s)$:
\be\lab{casbos}
 \Delta^b[m,\a] = {\sum_{n=1}^{\infty}}'\sqrt{(n\pm \a)^2 + m^2}.
\ee
Similarly, $\Delta^f[m,\a]$ is the Casimir energy of a twisted complex
fermion of mass $m$:
\be\lab{casfer}
 \Delta^f[m,\a] = {\sum_{n=1}^{\infty}}'\sqrt{(n-\half\pm \a)^2 + m^2}.
\ee
The primes denote a proper regularization of the sums which we 
give the precise definitions in the App. C. 
We note the following relation 
between the bosonic and fermionic zero-point energies:
\be\lab{casrel}
\Delta^f[m,\a] = \half\Delta^b[2m,2\a]-\Delta^b[m,\a],
\ee 
which is simply derived by dividing the sum in (\ref{casfer}) 
into odd and even parts. 

In order to demonstrate the modular properties of (\ref{pf1}) 
it is useful to introduce the parameter,
\be\lab{tau}
\t=\t_1+i\t_2 \equiv\l+i\frac{\b}{2\pi\a' p^+},
\ee
in terms of which (\ref{pf1}) becomes,
\be\lab{pf2}
\cz =-\frac{\b}{2\pi\a'}\int \frac{d^2\t}{\t_2^2} e^{\frac{-\b^2}{2\pi\a'\tau_2}}Z(\t)
\ee
where $Z(\t)$ is the trace in (\ref{pf1}). The $\tau$ integration 
is over the infinite strip $\tau_2\ge 0$, $1/2>\tau_1>-1/2$. 

The use of the Lagrange multiplier $\l$ in (\ref{pf1}) 
decouples the contributions from the excitations 
in different transverse directions. As usual, the fermionic and bosonic 
contributions can also be computed separately. Therefore the trace $Z$ can 
simply be evaluated as in the following product form:   
\be\lab{pf3}
Z(\t)=|q|^{H_0+\frac13}\left(\frac{|\eta(\t)|^{-2}}{\tau_2^{1/2}}\ri)^4
\prod_{n\in{\mathbf Z}} \frac{\le|1+e^{-2\pi\tau_2w_n^{++}+2\pi i n \tau_1}\ri|^4  
\le|1+e^{-2\pi\tau_2w_n^{+-}+2\pi i n \tau_1}\ri|^4}
{\le|1-e^{-2\pi\tau_2w_n^{y}+2\pi i n \tau_1}\ri|^2  
\le|1-e^{-2\pi\tau_2w_n^{z}+2\pi i n \tau_1}\ri|^2}.
\ee
The various contributions in this expression need explanation.
The second piece that involves the Dedekind eta function is the 
standard torus partition function of bosonic string excitations in
four flat directions. The eta function,
\be\lab{eta}
\eta(\t) = q^{\frac{1}{24}}\prod_{n=1}^{\infty}(1-q^n),
\ee
includes the sum over excitations of frequency modes $n\geq 1$ 
and the zero-point energy, $\Delta^b[0,0] = -1/12$.
The $\tau_2^{1/2}$ term comes from the Gaussian integration 
over the center of mass momentum in the flat directions.          
The exponential piece in front of (\ref{pf3}) arises 
from the zero-point energy of the system on the cylinder. 
$H_0$ is defined in (\ref{casimir}). We subtracted the piece 
that comes from the flat directions because they 
are already included in the definition of the $\eta$ function.
The numerator of the big product includes contributions 
from all of the fermionic excitations of mode $n\ge 0$ and finally 
the denominator of the product involves contributions of bosonic string 
excitations along the massive directions 5,6,7 and 8. We remark that 
the mode $n=0$ in the massive directions corresponds 
to the oscillatory motion of the center of mass of the string 
in the gravitational potential well along these massive 
transverse directions. We also note that (\ref{pf3}) reduces 
to the partition function that is computed 
in \cite{PandoZayas:2003jr} in the limit $\gb\to 0$, as it should.

\section{The UV asymptotics}
\setcounter{equation}{0}

The thermal partition function of a single-string in our pp-wave background 
is given by the eqs. (\ref{pf2}) and (\ref{pf3}). In this section, we study the
UV asymptotics of the partition function. The UV regime corresponds to $p^+\to\infty$ 
hence $\tau_2\to 0$ by (\ref{tau}). It is easy to see that the only possible 
divergence of $Z(\tau)$ (\ref{pf3}) in this limit, is at $\tau_1=0$. 
Thus, in the following we set $\tau_1=0$ and study the limit $\tau_2\to 0$ 
in order to derive the Hagedorn temperature above of which 
(\ref{pf2}) becomes ill-defined.

As the products in (\ref{pf3}) diverge in this limit, one should 
utilize the modular transformation properties of the various pieces 
in $Z(\tau)$. The modular S-transformation of the eta function is given by,
\be\lab{etaS}
\eta\le(i\tau_2\ri) = \tau_2^{-\half}\eta(i\tau_2^{-1}).
\ee
The infinite products in (\ref{pf3}) cannot be directly expressed in 
terms of well-known modular functions, hence their behavior under the S-transformation 
is less clear. Luckily, for $\tau_1=0$, they can be expressed in terms of
the {\em generalized modular functions} first introduced in \cite{Bergman:2002hv} and
further generalized by Takayanagi in \cite{Takayanagi:2002pi}. Takayanagi defines the following 
generalized Dedekind function, 
\be\lab{taka}
Z_{\d,\a}^{(m)}\le(\t\ri)=e^{4\pi\t_2\Delta^b[m,\a]}\prod_{n\in{\mathbf Z}}
\le|1-e^{-2\pi\tau_2\sqrt{m^2+(n+\a)^2}+2\pi i\tau_1(n+\a)+2\pi i \d}\ri|^2.
\ee
This function has nice transformation properties under the modular 
transformations. Its behavior under the S-transformation is derived by
Poisson re-summation in \cite{Takayanagi:2002pi} and it is shown that,
\be\lab{takaS}
Z_{\d,\a}^{(m)}(\t)=Z_{\a,-\d}^{(m|\t|)}(-\frac{1}{\t}).
\ee

It is not hard to see that, for $\tau_1=0$, the denominator of (\ref{pf3}) 
together with the corresponding zero point energy terms in the exponential 
can directly be expressed in terms of (\ref{taka}). The same is true for the 
numerator as well if one uses $(1+e^x) = (1-e^{2x})/(1-e^x)$. Using also 
the relation (\ref{casrel}), all in all one obtains,
\be\lab{pf4}
Z(i\tau_2) = \frac{|q|^{2(\Delta^b[\frac{4\mb}{3},0]
+\Delta^b[\frac{2\mb}{3},2\a])}{Z_{0,0}^{(2\mb/3)}(2i\t_2)}^2
{Z_{0,\a}^{(\mb/3)}(2i\t_2)}^2}
{\t_2^2|\eta(i\t_2)|^8Z_{0,\a}^{(\mb)}(i\tau_2)
{Z_{0,\a}^{(\mb/3)}(i\tau_2)}^3 
{Z_{0,0}^{(2\mb/3)}(i\tau_2)}^2},
\ee    
where $\a=\mb\gb/2$. Now it is a simple exercise to work out the 
UV asymptotics of (\ref{pf4}) if one uses the asymptotics 
of the Dedekind and generalized Dedekind functions as $\t_2\to 0$. 
These can easily be found by employing the modular transformation 
properties (\ref{etaS}) and (\ref{takaS}). Thus as $\t_2\to 0$ one obtains, 
\be\lab{asymp}
\eta(i\tau_2)\to\t_2^{-\half}e^{-\frac{2\pi}{\t_2}\frac{1}{24}},
\qquad Z_{\d,\a}^{(m)}(i\t_2)\to 
e^{\frac{4\pi}{\t_2}\Delta^b[m\t_2,-\d]}.
\ee
Substituting these asypmtotics in (\ref{pf2}) we find that, 
for temperatures larger than a $\b_H$ the thermal partition diverges
and the Hagedorn temperature is determined by the following formula:
\be\lab{hage}
-\le(\frac{\b_H}{2\pi\sqrt{\a'}}\ri)^2
-2\Delta^b[\frac{m_0\b_H}{2\pi},0]-2\Delta^b[\frac{m_0\b_H}{6\pi},0]
+4\Delta^f[\frac{m_0\b_H}{3\pi},0]+4\Delta^f[\frac{m_0\b_H}{6\pi},0]+\frac13=0.
\ee
In the derivation, we further used the relation (\ref{casrel}) and 
$\mb\t_2 = m_0\b/2\pi$.      
In fact the precise value of $T_H$ depends on the way one approaches the singular 
$\tau=0$ \cite{Greene:2002cd}. This fact shows up as dependence of $T_H$ 
on the fixed ratio $\theta =\t_1/\t_2$. However, the minimum of $T_H$ 
as a function of this ratio is at $\theta=0$, which 
coincides with \ref{hage}. Therefore, \ref{hage} defines the absolute temperature that signals 
an irregular behavior in the thermal partition function. In either case, 
$T_H$ stays unaltered under the $\g$-deformation. 

Another important subtlety in the expression (\ref{hage}) 
is the question that concerns the regularization of the Casimir energies.
As pointed out 
in \cite{Bigazzi:2003jk}, in fact, the sum over the Casimir energies that appear in 
(\ref{hage}) is finite without regularization, due to the underlying supersymmetry in 
the theory. Therefore \cite{Bigazzi:2003jk} argues that 
regularizing separate Casimir energies as in the App. C unnecessarily subtracts 
a finite term from this expression. Using that prescription, one instead obtains 
(\ref{hage}) where the sum over the regularized Casimir energies, $E_0'$, is replaced
by the unregulated sum $E_0= E_0'+\delta_0$. For our geometry, the difference is 
$\d_0 = (32/9 \log(2)- 4\log(3))\mb^2$. This additional term, alters the Hagedorn behavior 
in the limit $m_0\to\infty$. For our purposes in this paper, 
it suffices to note that, this modification 
does not alter our conclusion: $T_H$ stays unaltered under the $\g$-deformation
\footnote{We used the regularized sum in App. C in order to use the nice modular properties 
of the function (\ref{taka}). Shifting $H_0$ by $\d_0$ in (\ref{pf3}) 
gives the result with $E_0$ above, without the need to alter our discussion on the modular properties
of the various functions above. 
We thank A. Cotrone and F. Bigazzi for communication on these issues.}.           

It looks surprising that we obtained exactly the same Hagedorn temperature 
as the one for the pp-wave of the original MN background \cite{PandoZayas:2003jr}. 
We note that 
the parameter $\gb$ enters the spectra in a non-trivial way, 
therefore the thermal partition function in the 
${\rm MN}_{\g}$ pp-wave depends on $\gb$ non-trivially. 
It is easy to trace back the mathematical reason 
for the unexpected independence of $T_H$ on $\gb$: 
For $\t_1=0$, the dependence of $Z(\t)$ on $\gb$ is confined
in the coefficient $\a$ in (\ref{pf4}) that only appears inside 
the generalized Dedekind functions and the exponential in front.  
As under the S-transformation (\ref{takaS}),
the coefficients $\d$ and $\a$ are interchanged, the UV asymptotic of 
the generalized Dedekind function becomes independent of $\gb$. 
Similarly the dependence on $\a$ in the exponential in (\ref{pf4}) 
disappears as $\t_2\to 0$, \ie $q\to 1$. However, 
the dependence on $\gb$ is manifest in other regions of the moduli space $\t$. 
We discuss a simpler physical explanation in the next section.  
Apparently, the independence of the parameter $\a$ of 
the UV asymptotics of (\ref{taka}) was first noticed in \cite{Greene:2002cd}, 
in a similar 
computation concerning the maximally symmetric pp-wave background.  
In that paper, this parameter had no physical significance but kept just 
as a dummy index in the functions, 
in order to study their modular transformations. 
In the maximally symmetric background, 
it does not appear in the energy spectrum of the string excitations.

In order to understand the nature of the Hagedorn temperature, one needs to look at  
the free-energy of an ideal gas of strings in this background as in 
\cite{PandoZayas:2003jr}\cite{Brower:2002zx}. 
Although we did not carry out this computation 
(because our main focus of discussion is rather different in this paper), 
the fact that the dependence on $\g$ only enters as $n\to n+ \g$ 
we strongly believe that the free-energy stays finite at $T_H$ also in our case. 
Therefore it is reasonable to conjecture that 
the Hagedorn behavior signals a phase transition also in the $\g$ deformed background.    
 
\section{Discussion and outlook}

In this paper, we investigated the idea that a quantity that is independent of the 
deformation parameter $\g$ in the \mng background is likely to give reliable information 
on the \sym in the IR. One obvious loophole is that, perhaps there exist quantities 
which are independent of the deformation not because they do not receive 
contributions from the KK-sector but because their dependence on the dipole deformation 
vanishes either because of some symmetry reasons or in some particular limits 
that render the effects of the 
dipole deformation negligible. It is quite important to make this point 
as rigorous as possible. 

Indeed, our construction in this paper illustrates this possibility. 
Although the pp-wave that we constructed is conjectured to describe the dynamics 
of KK-hadrons, there exist quantities in the thermal domain that do not exhibit 
dependence on the dipole parameter $\g$. Thus, our example makes it clear that 
independence of a field theory quantity of the deformation is only a 
{\em necessary but not sufficient} condition for the universality. 
However we beleive that one can argue for true universality by supplying this 
necessary condition with additional arguments. For example, 
we beleive that certain quantities 
computed in \cite{Gursoy:2005cn} fall into this class. 
   
As an illustration of a quantity that is independent of $\g$, 
we computed the Hagedorn temperature of the strings 
in a pp-wave that is obtained from the \mng background of 
\cite{Gursoy:2005cn}. The reason that this quantity turns out to be 
independent of $\g$ can be understood as follows. The Hagedorn temperature is given 
by the density of states in the UV region. In this region, the string excitation number 
effectively becomes a continious parameter and the density of states is given by
$\r(w) = (dw(n)/dn)^{-1}$. By looking at our formulas for the spectrum in the $\g$ 
deformed pp-wave background, we notice that the whole effect of the deformation 
is the shifts $n\to n +\g$. Thus $T_H$ is expected to be independent of $\g$.
This serves as a simple illustration of how to construct 
a non-universal quantity, yet independent of the deformation parameter $\g$.  

There are several other directions that one would like to explore. It would be very 
interesting to obtain examples of true ``universal'' quantities from the \mng 
background which preferably admit an easier interpretation from the field theory 
point of view. For a quantity $Q$ that is independent of $\g$, one can argue 
for the universality as follows. One would like to say that in the limit 
$\gb\to\infty$, $Q$ stays the same but the remnants of the KK-sector is swept 
out of the spectrum. This is not quite true because as seen from the figure 1, 
whatever the value of $\gb$ is, the region of the KK-spectra near the 
minimum of the $w_-$ branch never really decouples. However, in this limit 
the contribution from this region carry extremely high excitation number $n$. 
Thus, if one constructs $Q$ such that it receives contributions mainly from the
lowest lying excitations, then one can argue that this quantity carries reliable 
information on the pure SYM. Here the excitation number $n$ should be dual to a quantum 
number carried by the glueballs in the large $N$ limit. 
The Hagedorn temperature is not of this 
type because in the torus partition function, the contribution from the lowest 
and highest modes are all mixed up through the modular transformation properties.    
  
Secondly, one may apply our methods to the case of the pp-wave that 
is obtained from the KS background in \cite{Gimon:2002nr}. We expect that the 
$\g$ deformed KS background \cite{Lunin:2005jy} also admits a pp-wave similar 
to ours, and that $T_H$ is again independent of $\g$. This should be straightforward 
to work out, albeit technically more involved than our case. As the field theory 
that corresponds to the KS pp-wave is better understood \cite{Gimon:2002nr}, 
this exercise would help clarifying our discussion above. It should also be very 
interesting to carry out a similar analysis in the backgrounds that are dual 
to the $\g$ deformation of the ${\cal N}=1$ theories with flavors 
\cite{Casero:2006pt}. 

Another interesting direction would be to study the BMN type states dual to 
the pp-wave in question. A similar study was done in \cite{Niarchos:2002fc}. 
The BMN states should involve non-trivial dependence on the parameter $\g$. The
way that this dependence arises is suggested by the fact that our pp-wave spectrum 
is the spectrum of fields with twisted b.c. on the cylinder, \ie 
$\phi(\s+\pi) = e^{2\pi i\g}\phi(\s)$ (see \cite{Landsteiner:2006qb} 
for a discussion on this point). A similar one-parameter 
generalization of the BMN states is found in \cite{Takayanagi:2002hv}. 
After construction of these states, it will be 
interesting to study, for example, 
their supersymmetry properties \cite{Gursoy:2002yy}, \cite{Beisert:2002tn}.  
 
A question that is independent of the above concerns is the stringy corrections to the 
\mng background. We believe that our pp-wave solution is an exact solution to string 
theory including the $\a'$ corrections by the argument of \cite{Amati:1988ww}, 
\cite{Horowitz:1989bv}. However $g_s$ corrections are much harder to study than 
the case of the flat space or the maximally symmetric pp-wave of \cite{Blau:2001ne}. 
In the latter cases the $g_s$ corrections to the light-cone Hamiltonian follow
from symmetry restrictions \cite{Green:1983hw}, \cite{Spradlin:2002ar} 
that are less restrictive here. Apart from this discussion, the fact that $\g$ 
is directly proportional to the value of the axion in the pp-wave 
geometry may give us a hint about the stringy corrections 
to the \mng background: perhaps these corrections should 
promote the parameter $\g$ to a dynamical variable. 

One can also ask whether or not our pp-wave geometry 
can directly be obtained as a deformation of the MN pp-wave \cite{Gimon:2002nr}
\footnote{We thank Elias Kiritsis for attracting our attention to this possibility.}. 
The $SL(3,R)$ transformations used in this paper when directly applied to a 
pp-wave geometry generally does not result in another pp-wave geometry. However, 
a class of deformations of a pp-wave that again give rise to pp-waves are 
discussed \eg in \cite{D'Appollonio:2003dr}. It would be quite interesting to 
explore this point further.          

Finally, we would like to point out an interesting application of our results to
the $\b$-deformed ${\cal N}=4$ SYM theory. In \cite{Lunin:2005jy}, a pp-wave geometry 
is constructed from the gravity dual of this field theory. By comparison of the 
spectra of that pp-wave to the pp-wave considered here, one realizes that our arguments 
about the independence of $T_H$ of the deformation parameter possibly holds for that 
case as well. This fact has immediate consequences for the thermal properties of 
the corresponding field theory. It implies that, in the limit of large $R$-charge two 
different field theories, a superconformal \sym theory and the ${\cal N}=4$ SYM theory 
has exactly the same Hagedorn behavior. It should be interesting to pursue this 
observation further in the light of \cite{Aharony:2003sx}. 

\section{Acknowledgments}

It is a pleasure to thank Chang-hyun Ahn, Francesco Bigazzi, 
Roberto Casero, Aldo Cotrone, Mariana Gra\~na, Clifford Johnson, 
Elias Kiritsis, Francesco Nitti, Carlos N\'u\~nez and Angel Paredes-Galan
for useful discussions. 
This work was partially supported by ANR grant NT05-1-41861,
INTAS grant, 03-51-6346,
RTN contracts MRTN-CT-2004-005104 and MRTN-CT-2004-503369,
CNRS PICS 2530 and 3059 and by a European Excellence Grant,
MEXT-CT-2003-509661.

\newpage
    
\appendix

\section{Details of the background}

In this appendix, we present some details of the \mng background. 
For completeness let us first give expressions for the quantities that appear in 
the MN background. The one-forms that define the fibration of the $S^2$ on the $S^3$ 
in (\ref{mn}) are, 
\be\lab{gf}
A^1\,=\,-a(r) d\theta\,,
\,\,\,\,\,\,\,\,\,
A^2\,=\,a(r) \sin\theta d\varphi\,,
\,\,\,\,\,\,\,\,\,
A^3\,=\,- \cos\theta d\varphi\,,
\ee
and the the left invariant $SU(2)$ one-forms are given by,  
\bea\lab{su2}
w_1&=& \cos\psi d\tilde\theta\,+\,\sin\psi\sin\tilde\theta
d\tilde\varphi\,\,,\rc
w_2&=&-\sin\psi d\tilde\theta\,+\,\cos\psi\sin\tilde\theta
d\tilde\varphi\,\,,\rc
w_3&=&d\psi\,+\,\cos\tilde\theta d\tilde\varphi\,\,.
\eea
Ranges of the three angles are $0\leq\tilde\varphi< 2\pi$, $0\leq\tilde\theta\leq\pi$ and
$0\leq\psi< 4\pi$. 
Various functions that appear in (\ref{mn}) and the dilaton of the MN geometry is 
given by,
\bea
a(r)&=&{2r\over \sinh 2r}\,\,,\rc
e^{2h}&=&r\coth 2r\,-\,{r^2\over \sinh^2 2r}\,-\,
{1\over 4}\,\,,\rc
e^{-2\phi}&=&e^{-2\phi_0}{2e^h\over \sinh 2r}\,\,,
\label{MNsol}
\eea
The background includes an RR two-form $C^{(2)}_0$, that is given by
\bea
C^{(2)}_0&=&{1\over 4m_0^2}\,\Big[\,\psi\,(\,\sin\theta d\theta\wedge d\varphi\,-\,
\sin\tilde\theta d\tilde\theta\wedge d\tilde\varphi\,)
\,-\,\cos\theta\cos\tilde\theta d\varphi\wedge d\tilde\varphi\,-\rc
&&-a\,(\,d\theta\wedge w^1\,-\,\sin\theta d\varphi\wedge w^2\,)\,\Big]\,\,.
\label{RR2}
\eea
This concludes the definitions concerning the MN geometry. 

The angular part of the \mng metric in (\ref{metg}) reads,
\bea\lab{dw5}
d\Omega_5^2 &=& D_1\, d\q^2+D_2\, d\qt^2 + D_3\, d\j^2 +D_4\, d\f^2+ D_5\, d\ft^2 
+ E_1\, d\q\, d\qt+ E_2\, d\q\, d\j\\
{}&&+E_3\, d\qt\, d\j +  E_4\, d\q\, d\ft 
+ E_5\, d\qt\, d\f + E_6\, d\j\,d\f + E_7\, d\j\, d\ft + E_8\, d\f\, d\ft,\nn
\eea
where,
\bea\lab{metf}
F &=&  \frac{e^{\phi}}{4m_0^2} \sqrt{f-g^2},\,\, 
D_1 = f a^2 \sin^2{\j}\,\sin^2{\qt},\nn\\
D_2 &=& a^2 \sin^2{\q}\, \sin^2{\j}, \,\,
D_3 = f \cos^2{\qt} + \cos^2{\q} + 2g \cos{\q}\,\cos{\qt},\nn\\
E_1 &=& 2a^2\, g\, \sin^2{\j}\,\sin{\q}\, \sin{\qt},\,\,
E_2 = 2a\, \sin{\j}\, \sin{\qt} (f\, \cos{\qt} + g\, \cos{\q}),\nn\\
E_3 &=& 2a \sin{\j}\,\sin{\q}(\cos{\q} + g\, \cos{\qt}),\,\,   
D_4 = f(f-g^2),\, D_5 = f-g^2,\\
E_4 &=& 2a \sin{\j} \sin{\qt} (f-g^2), \,\, 
E_5 = 2a \sin{\j}\, \sin{\q} (f-g^2), \nn\\ 
E_6 &=& 2\cos{\q}(f-g^2), \,\, E_7 = 2\cos{\qt}(f-g^2), \,\, 
E_8 = 2f(f-g^2),\nn\\
f &=& 4 e^{2h} \sin^2{\q} + \cos^2{\q} + a^2\, \sin^2{\q}, 
\,\, g= a\,\sin{\q}\,\sin{\qt}\,\cos{\j}-\cos{\q}\,\cos{\qt}.\nn
\eea

The one-forms that appear in the RR-forms and the NS-form of \mng in 
eqs. (\ref{RR2g}), etc are,  
\be\lab{A1}
{\cal A}_1 = (f-g^2)^{-1}\Bigg\{(a \sin\tilde{\theta} \sin\psi g)d\q+ (a
\sin\theta \sin\psi) d\tht + (\cos\theta + g
\cos\tilde{\theta}) d\psi\Bigg\}.
\ee
and 
\be\lab{A2}
{\cal A}_2  = (f-g^2)^{-1}\Bigg\{f(a \sin\psi \sin\tilde\theta) d\q + g a (\sin\psi
\sin\theta) d\tht + (f \cos\tilde\theta + g \cos\theta) d\psi\Bigg\}.
\ee

All of these expressions involve many simplifications with respect to the form 
that was originally presented in \cite{Gursoy:2005cn}. 

\section{The coordinate transformation}

The coordinate transformation (\ref{cov}) of section 3.1 is,  
\bea\lab{cov2}
e^{\frac{i}{2}(\oj + \of)}\cos{\frac{\oq}{2}} &=& 
e^{\frac{i}{2}(\j + \ft-\f)}\cos{\frac{\q}{2}}\cos{\frac{\qt}{2}} +     
e^{-\frac{i}{2}(\j - \ft-\f)}\sin{\frac{\q}{2}}\sin{\frac{\qt}{2}},\nn\\
e^{\frac{i}{2}(\oj - \of)}\sin{\frac{\oq}{2}} &=& 
e^{\frac{i}{2}(\j - \ft-\f)}\cos{\frac{\q}{2}}\sin{\frac{\qt}{2}} -     
e^{-\frac{i}{2}(\j + \ft-\f)}\sin{\frac{\q}{2}}\cos{\frac{\qt}{2}}. 
\eea
There are three linearly independent equations contained in (\ref{cov2}). 
We also need the expansion of these coordinate transformations in the first few 
powers in $R$: 
\bea\lab{covs} \cos{\oq} &=& \cos{\q} + \frac{\qt}{R} \sin{\q}\cos(\j-\f)
-\frac{\qt^2}{2R^2}\cos{\q}+ {\cal O}(1/R^3), \nn\\
\sin{\oq} &=& = \sin{\q} - \frac{\qt}{R} \cos{\q}\cos(\j-\f) 
+\frac{\qt^2}{4R^2} \frac{\cos{2\q}-\cos{2(\j-\f)}}{\sin{\q}} 
+ {\cal O}(1/R^3), \nn\\
\cos{\oj} &=& -1 + \frac{\qt^2}{2R^2}\frac{\sin^2(\j-\f)}{\sin^2{\q}} + 
 {\cal O}(1/R^3), \\
\sin{\oj} &=& \frac{\qt}{R} \frac{\sin(\j-\f)}{\sin{\q}}+{\cal O}(1/R^3).\nn
\eea
For the Penrose limit of the various n-forms in the geometry, 
in fact, one needs more than the first terms of the line elements in (\ref{dcovs}):
\bea\lab{dcovs2}
d\oq &=& d\q-\frac1R\le(d\qt\cos(\j-\f)
-\qt\sin(\j-\f)(d\j-d\f)\ri)+{\cal O}(1/R^2),\\
d\oj &=& -\frac{1}{R\sin\q}\le(d\qt\sin(\j-\f)
+\qt\cos(\j-\f)(d\j-d\f)-\qt\cot\q\sin(\j-\f)d\q\ri)+{\cal O}(1/R^2).\nn
\eea

\section{The regularized Casimir energies}

The regularization of the Casimir energies in (\ref{casbos}) and (\ref{casfer}) are 
discussed in \cite{PandoZayas:2002hh} and \cite{Takayanagi:2002pi} 
and most extensively in \cite{Bigazzi:2003jk}. 
Throughout the paper we use the definition, 
\be\lab{casreg}
{\sum_{n=1}^{\infty}}'\sqrt{(n+a)^2+m^2} = -\frac{1}{2\pi^2}
\sum_{p=1}^{\infty}\int_0^{\infty}ds e^{-p^2 s -\frac{\pi^2m^2}{s}}\cos(2\pi pa).
\ee
The fermionic sum can similarly be defined by making the substitution $a\to a-1/2$ 
above.

\end{document}